\documentclass[aps,pra,twocolumn,floatfix,showpacs]{revtex4}

\usepackage{graphicx}

\begin{document}

\title{Evaluation of the $^{52}$Cr-$^{52}$Cr interaction via spin-flip
scatterings}

\author{Y. Z. He}
\author{Z. F. Chen}
\author{Z. B. Li}
\author{C. G. Bao}
\email{stsbcg@mail.sysu.edu.cn}
\thanks{The corresponding author}
\affiliation{Department of Physics, and State Key Laboratory of
Optoelectronic Materials and Technologies, Sun Yat-Sen University,
Guangzhou, 510275, P.R. China}


\begin{abstract}
In order to evaluate $g_0$, the interaction strength of a pair of
$^{52}$Cr atoms with total spin $S=0$, a specially designed s-wave
scattering of the pair has been studied theoretically. Both the
incident atom and the target atom trapped by a harmonic potential
are polarized previously but in reverse directions. Due to
spin-flip, the outgoing atom may have spin component $\mu$ ranging
from -3 to 3. The outgoing channels are classified by $\mu$. The
effect of $g_{0}$ on the scattering amplitudes of each of these $\mu
-$channels has been predicted.
\end{abstract}

\pacs{03.75.Mn, 34.20.Cf, 34.10.+x}

\maketitle

The chromium atoms $^{52}$Cr are special due to having a larger spin
$F=3$ and a larger magnetic moment $6\mu _{B}$. Since the
experimental realization of the chromium condensation
\cite{r_GA2005}, the interest in this field has been increasing. The
$^{52}$Cr condensate is a new kind of matter aggregation having the
magnetic dipole-dipole interaction $V_{dd}$ more than twenty times
stronger than that of the alkalis family. A direct consequence of
$V_{dd}$ is the coupling of the spatial and spin degrees of freedom
so that the conversion of spin angular momentum into orbital angular
momentum can be realized. Thereby new physical phenomena (say,
creation of vortex) might appear. In addition to $V_{dd}$, the
atom-atom interaction depends strongly on the total spin $S$ of the
pair, and is in general written as $ V_{12}=\delta
(\mathbf{r}_{1}-\mathbf{r}_{2})\sum_{S}g_{S}\mathfrak{P}
^{S}+V_{dd}$, where $g_{S}$ is the strength and $\mathfrak{P}^{S}$
is the projection operator of the $S-$spin-channel ($S=0,\ 2,\ 4$
and 6). Up to now, $g_{2},\ g_{4}$, and $g_{6}$ have been
determined, but $g_{0}$ has not yet
\cite{r_WJ2005,r_SJ2005,r_DRB2006,r_GA2006}. However, many features
of the condensate depend on this parameter (say, the phase-diagram
of the ground state depends strongly on $g_{0}$
\cite{r_DRB2006,r_MH2007,r_US2008,r_VP2007}. The dependence is also
explicit in spin-evolution \cite{r_SL2006}).  Therefore, the
determination of $g_{0}$ is important for a thorough and clear
description of the condensates. This paper is one along this line.

The effect of $g_{0}$ can be exposed via two-body scatterings. We
study the scattering of a very slowly incident $^{52}$Cr atom by
another $^{52}$Cr atom trapped previously in a potential $U(r)$.
The energy of collision is so low that the trapped atom is not able
to be excited, and only the s-wave of the incoming one is involved.
The scattering should be so designed that related observables are
sensitive to $g_{0}$. The following points are proposed to meet the
goal.

(i) When $M_S$, the Z-component of the total spin $S$ of the pair,
is non-zero, $g_{0}$ plays no role. Therefore, the two-body system
must be designed to have $M_S=0$. We therefore assume that the
spin-state $\chi _{\mu }$ of the incident particle has $\mu =3$,
while that of the target particle in the trap has $\mu =-3$. This is
different from the design of \cite{r_WJ2005}, where both particles
have ¦Ì=-3. With their design, they have successfully determined the
values of $g_{2}$, $g_{4}$, and $g_{6}$. However, $g_{0}$ can not be
thereby determined.

(ii) In general, for a scattering by a potential $U(r)$, the
cross-section depends on the parameters of the potential very
sensitively. For an example, we study the case of a finite harmonic
potential $U_{p}(r)=\frac{1}{2}m\omega ^{2}(r^{2}-r_{0}^{2})$ if
$r\leq r_{0}$, or $U_{p}(r)=0$ if $r>r_{0}$. The parameter $r_{0}$
measures the width and depth of the potential. In what follows
$\hbar \omega $ and $\lambda \equiv \sqrt{\hbar /m\omega }$ are used
as units of energy and length, respectively. When a single $^{52}$Cr
atom with an incident kinetic energy $k^{2}/2$ is scattered by this
potential, the $s-$wave cross-section $\sigma _{U}$ against the
parameter $r_{0}$ is shown in Fig.1, where $k=0.1$. One can see, if
$r_{0}$ is given in a narrow domain around 2.512, $\sigma _{U}$
would be much larger. In the follows two kinds of potentials will be
used, and their parameters will be selected in the narrow domains
that lead to large $\sigma _{U}$.
\begin{figure}[tbp]
\centering \resizebox{0.95\columnwidth}{!}{
\includegraphics{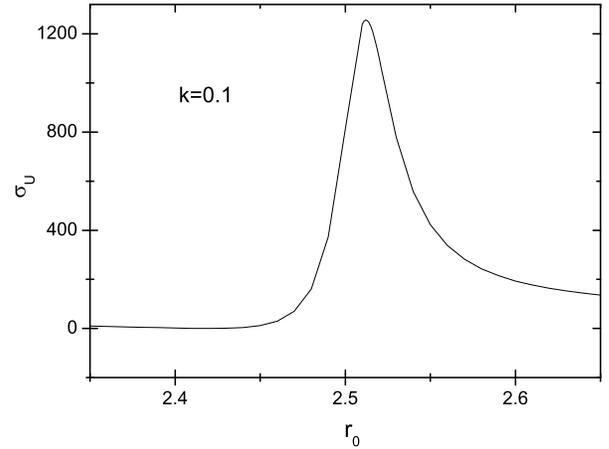} } \caption{The $s-$wave cross section
$\sigma _{U}$ of a particle scattered by a finite harmonic potential
$U_{p}$. $r_{0}$ is the width of the potential, and $k=0.1$ is
assumed. The units of $ \sigma _{U}$, $r_{0}$, and $k$ in this paper
are $\lambda ^{2}$, $\lambda $, and $\lambda ^{-1}$, respectively.
$\lambda =\sqrt{\hbar /m\omega }$ (say, if $\omega =300\times 2\pi$,
then $\lambda =806nm$ for the $^{52}$Cr atom). }
\end{figure}

(iii) The dipole-dipole interaction is given by
\begin{equation}
 V_{dd}
 =\frac{C_{d}}{r^{3}}
  (\mathbf{F}_{1}\cdot\mathbf{F}_{2}
  -3\frac{ (\mathbf{F}_{1}\cdot\mathbf{r})
           (\mathbf{F}_{2}\cdot\mathbf{r})}{r^{2}})
\end{equation}
where $\mathbf{F}_{i}$ is the operator of the spin of the $i-$th
atom, and $\mathbf{r=r}_{2}\mathbf{-r}_{1}$. It can be proved that
$V_{dd}$ must cause an additional spatial relative rotation of the
two particles in d-wave as shown in the appendix. However, if the
total orbital angular momentum $L$ can be kept zero (say, due to a
constraint in energy), the effect of $V_{dd}$ would be strictly
suppressed, and therefore can be neglected. The neglect of $V_{dd}$
would greatly facilitate the calculation shown below. This is
different from the design of \cite{r_WJ2005}, where a magnetic field
is applied. It is reminded that, if higher partial waves are
introduced by $V_{dd}$, they are accompanied by large-$S$ components
due to the conservation of the total angular momentum. Among these
components those with a large-$M_{S}$ will lead to a decrease of
energy due to the Zeeman effect. Therefore, the increase of rotation
energy can be partially canceled by the additional Zeeman energy.
Consequently, the magnetic field would help $V_{dd}$ to cause a
stronger mixing to include higher partial waves, and the effect of
$V_{dd}$ thereby becomes remarkable. This is the case of
\cite{r_WJ2005}.

With the above consideration, for numerical calculations, two
potentials are introduced. One is the finite harmonic well
$U_{p}(r)$ with the parameter $r_{0}$ mentioned above, the other is
the Gaussian potential $U_{a}(r)=-4e^{-(r/a)^{2}}$ with the
parameter $a$. The propagation number of the incoming atom $k\leq
0.2$ so that the related kinetic energy $k^{2}/2$ is much smaller
than the energy gap of the ground state, and the maximal effective
classical angular momentum of the incident particle $\hbar k r_0$ is
smaller than $\hbar$. With these choices the target atom can not be
excited, and only the s-waves of both atoms are involved, and the
total orbital angular momentum $L$ remains to be zero as required.

When $V_{dd}$ is dropped, $S$ and $M_S$ are good quantum numbers.
Let the total wave function $\Psi$ be a composition of different
$S-$component,
\begin{equation}
 \Psi =\sum_{S}C_{S}\Psi _{S}
 \equiv \sum_{S}C_{S}\frac{1}{r_{1}r_{2}}
                \Phi _{S}(r_{1},r_{2})(12)_{SM_S}
\end{equation}
where $S$ is from 0 to 6 and $M_S=0$ as designed. Each $\Psi _{S}$
should satisfy the Schr\"{o}dinger equation, while the coefficients
$C_{S}$ are determined by how the incoming channel would be. Since
only s-waves are involved, the angular degrees of freedom are
irrelevant. Thus the Schr\"{o}dinger equation for $\Phi _{S}$ reads
\begin{eqnarray}
 [E \
  +\frac{d^{2}}{2dr_{1}^{2}}
  &&+\frac{d^{2}}{2dr_{2}^{2}}
  -U(r_{1})
  -U(r_{2}) \nonumber \\
  &&-g_{S}\ \delta(\mathbf{r}_{1}-\mathbf{r}_{2})\ ]\
 \Phi _{S}(r_{1},r_{2})=0
\end{eqnarray}
Where the total energy $E=\varepsilon _{g}+k^{2}/2$, and
$\varepsilon _{g}$ is the ground state energy (negative) of the
target particle. Incidentally, $\varepsilon _{g}$ and the associated
wave function $\phi _{g}$ are easy to be obtained. In order to
facilitate numerical calculation, the $\delta -function$ is replaced
by $e^{-\beta r^{2}}/(\pi /\beta )^{3/2}$, where $\beta $ is chosen
to be a very large number to assure a very short range. We choose
$\beta =1/(0.06)^{2}$, and we found that the effect of the change of
$\beta $ in a reasonable domain is slight. When $S$ is even, $\Phi
_{S}$ must be symmetric with respect to the interchange of $r_{1}$
and $r_{2}$, whereas when $S$ is odd, it must be anti-symmetric.

For $k=0.1$ and $U=U_{p}$, we know from Fig.1 that the optimal
$r_{0}$ leading to the largest $\sigma _{U}$ is 2.512. Accordingly,
three cases with $r_{0}=$2.5, 2.512, and 2.53 are chosen to see the
effect of the variation of $r_{0}$ around its optimal value. We
define a domain (0,5) for both $r_{1}$ and $r_{2}$. Then, Eq.(3) is
solved numerically in the domain. The crucial point is the embedment
of boundary conditions. Obviously, we have $\Phi _{S}(0,r_{2})=\Phi
_{S}(r_{1},0)=0.$ Furthermore, it is reminded that, due to the
constraint in energy, if one atom is far away, the other one would
remain in the trap in the ground state (the outgoing of both atoms
is not possible). Therefore, when $r_{2}$ is sufficiently large,
$U(r_{2})=0$, $V_{12}=0$, and $\Phi _{S}(r_{1},r_{2})$ would tend to
$\phi _{g}(r_{1})(b_{S}\cos (kr_{2})+a_{S}\sin (kr_{2}))$, where the
second factor is the common form of s-waves. Both $b_{S}$ and
$a_{S}$ are real numbers, and the unitarity is thereby assured. When
$r_{2}=5$, the second factor is just a constant denoted by $h_{S}$.
Now we arrive at the second pair of boundary conditions, namely,
$\Phi _{S}(r_{1},5)=\phi _{g}(r_{1})h_{S}$ and $\Phi
_{S}(5,r_{2})=(-1)^{S}\phi _{g}(r_{2})h_{S}$. Where, $h_{S}$ can be
arbitrary given because the eventual results do not depend on it as
shown below. The factor $(-1)^{S}$ is needed to assure the correct
permutation symmetry.

Now, the values of $\Phi _{S}(r_{1},r_{2})$ at all the borders are
specified. Thus the problem is to solve an elliptic partial
differential equation under the Dirichlet condition. Numerical
solutions can therefore be obtained by using standard programs of
difference equations. For $S=2,\ 4$, and 6, $g_{S}$ have been known.
Using $\hbar \omega$ and $\lambda $ as units, $g_{2}=-0.134\times
10^{-3}\sqrt{\omega }$, $g_{4}=1.11\times 10^{-3}\sqrt{\omega }$,
and $g_{6}=2.14\times 10^{-3}\sqrt{\omega }$. For $S=0$, $g_{S}$
will be given at a number of values from $-2g_{6}$ to $2g_{6}$. When
$S$ is odd, $g_{S}$ is simply zero. Examples of $\Phi _{S}$ are
given in Fig.2. One can see in Fig.2a that the two atoms can be very
close to each other in the trap if $S$ is even, but can not if $S$
is odd as shown in 2b. Obviously, $\Phi _{S}(r_{1},r_{2})$ is
symmetric (anti-symmetric) in 2a (2b) as expected.
\begin{figure}[tbp]
\centering \resizebox{0.8\columnwidth}{!}{
\includegraphics{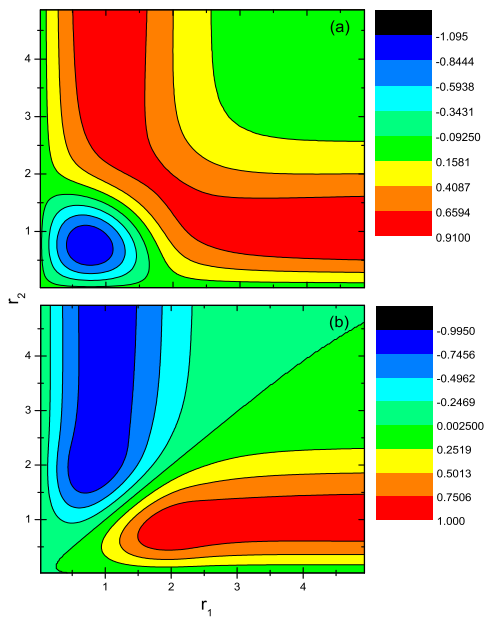} } \caption{Contour plots of
$\Phi _{S}(r_{1},r_{2})$\ with $S=6\ $(a) and $S=odd$ (b). The
potential is $U_{p}(r)$ with $r_{0}=2.5$, and $k=0.1$. $\omega
=300\times 2\pi$ is assumed (the same in the following figures). }
\end{figure}

Once $\Phi _{S}$ has been solved numerically, we can extract $b_{S}$
and $a_{S}$ from its asymptotic form. It is reminded that the
incoming particle has $\mu =3$, and the target particle has $\mu
=-3$. To meet this requirement, we choose $C_{S}=C_{3,3,\
3,-3}^{S,0}/(a_{S}-ib_{S})$. With this choice, in the asymptotic
region (say, $r_{1}$ is large while $r_{2}$ is small), the total
wave function appears as
\begin{eqnarray}
 \Psi
 &\rightarrow& \frac{1}{r_{1}r_{2}}\phi _{g}(r_{2})
  [\ \sin(kr_{1})\chi _{3}(1)\chi _{-3}(2) \nonumber \\
 &&  +\sum_{\mu }f_{\mu }\exp(ikr_{1})\chi _{\mu }(1)\chi _{-\mu }(2)\ ]
\end{eqnarray}
where
\begin{eqnarray}
f_{\mu }=\sum_{S} C_{3,3,\ 3,-3}^{S,0}\frac{b_{S}}{a_{S}-ib_{S}}
C_{3\mu ,\ 3,-\mu }^{S,0} \nonumber
\end{eqnarray}
where both even and odd $S$ are included in the summation. $f_{\mu}$
is the s-wave scattering amplitude of the $\mu -$channel (i.e., the
outgoing particle has spin component $\mu $, while the inner one has
$-\mu$ ). The asymptotic form with a large $r_{2}$ is similar. From
(4) we know that, in addition to the scattering by the potential,
spin-flips might occur due to the spin-dependent interaction
$V_{12}$. Incidentally, since $C_{S}$ contains the factor
$(a_{S}-ib_{S})^{-1}$ and both $a_{S}$ and $b_{S}$ are proportional
to $h_{S}$, $\Psi $ depends not at all on the set of constants
$h_{S}$ chosen before.

\begin{figure}[tbp]
\centering \resizebox{0.8\columnwidth}{!}{
\includegraphics{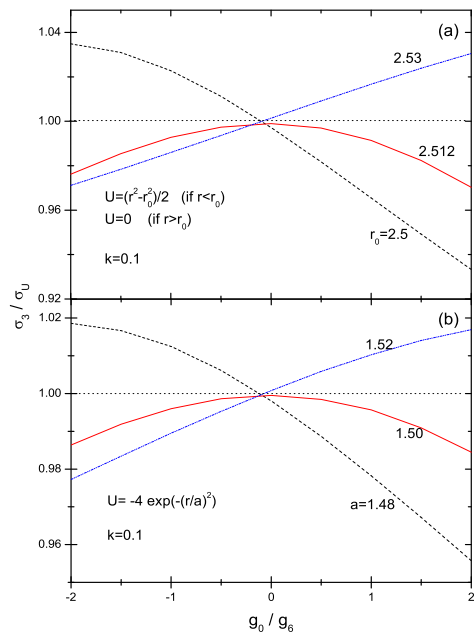} } \caption{(color online),
$\sigma _{3}/\sigma _{U}$ against $g_{0}/g_{6}$ with $k=0.1$. Finite
harmonic potential $U_{p}$ and Gaussian potential $U_{a}$ are used
in a and b, respectively. In 3a, the parameter $r_{0}$ is given at
2.5, 2.512, and 2.53 marked by the curves. The corresponding $\sigma
_{U}$ are 814, 1257(the optimal value), and 779, respectively. In
3b, the parameter $a$ is given at 1.48, 1.50, and 1.52. The
corresponding $\sigma _{U}$ are 978, 1257 (the optimal value), and
1003, respectively. }
\end{figure}

We are interested in how the s-wave cross section of a $\mu
-$channel $\sigma _{\mu }\equiv 4\pi |f_{\mu }|^{2}/k^{2}$ is
affected by $g_{0}$. It was found that $\sigma _{3}$ (elastic
channel without spin-flip) is much larger than other $\sigma _{\mu
}$. When the cross section of pure potential-scattering $\sigma
_{U}$ is considered as a unit, the ratio $\sigma _{3}/\sigma _{U}$
against $g_{0}$ is plotted in Fig.3. Four points are noted:

(i) The magnitude of $\sigma _{3}$ is in general close to $\sigma
_{U}$. In particular, if $g_{0}$\ is close to $g_{2}$, $\sigma _{3}$
would be very close to $\sigma _{U}$.

(ii) $\sigma _{3}$ depends sensitively on the potential parameter,
because $\sigma _{U}$ does. It implies that an appropriate choice of
parameters is crucial.

(iii) When the potential is narrower than the one leading to the
optimal $\sigma _{U}$, if $g_{0}$ is smaller than $g_{2}$, a more
negative $g_{0}$ would lead to a larger $\sigma _{3}$, while a more
positive $g_{0}$ lead to a smaller $\sigma _{3}$ (refer to the dash
curves of Fig.3). When the potential is broader, the effect of
$g_{0}$ is in reverse (refer to the dash-dot-dot curves). However,
when the parameter of the potential is optimized, a larger
$|g_{0}-g_{2}|$ would lead to a smaller $\sigma _{3}$ (refer to the
bold curves).

(iv) The above qualitative feature does not depend on the details of
the potential. This is explicit by comparing 3a and 3b, where
distinct potentials are used. Thus, for selecting an appropriate
potential, the crucial point is the understanding of the optimal
parameter.

Incidentally, in all the cases of Fig.3 the ground state is deeply
bound. For an example, when $U_p (r)$ has $r_0=2.5$, the ground
state energy is -1.6294, while the first excited state is about 1
higher. Thus the gap is greatly larger than the bombarding energy
$k^2/2\leq 0.02$.

\begin{figure}[tbp]
\centering \resizebox{0.8\columnwidth}{!}{
\includegraphics{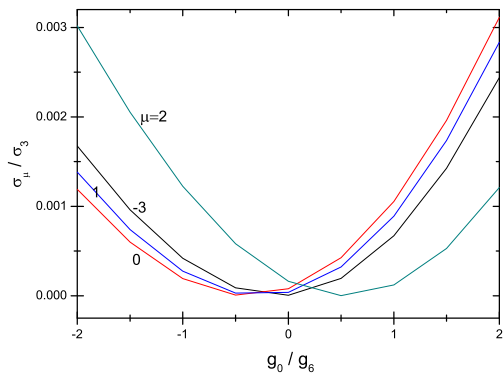} } \caption{(color online),
$\sigma _{\mu }/\sigma _{3}$ against $g_{0}/g_{6}$. The potential
$U_{p}$ with $r_{0}=2.512$ (the optimal value) is used, and $k=0.1$.
$\mu $ is marked by the curves. $\sigma _{-1}$ overlaps $\sigma
_{1}$, and $\sigma _{-2}$ overlaps $\sigma _{2}$. }
\end{figure}

In Fig.4 $\sigma _{\mu }/\sigma _{3}$ against $g_{0}$ is plotted.
Three points are noted:

(i) The cross-sections of spin-flip channels $\sigma _{\mu }$ are
remarkably smaller than that of the non-spin-flip channel.

(ii) It is reminded that the three $\Psi _{S}$ with odd $S$ are free
from atom-atom interaction, therefore they have the same
$b_{S}/(a_{S}-ib_{S})$. Since $\sum_{S}^{\prime }C_{3,3,\
3,-3}^{S,0}C_{3\mu ,\ 3,-\mu }^{S,0}=(\delta _{\mu ,3}-\delta _{\mu
,-3})/2$, where the summation covers only the three odd $S$, it is
straight forward to prove $f_{-\mu }=f_{\mu }$ if $\mu \neq \pm 3$.
Hence, the cases with $\mu =-2$ and -1 can not be seen in the
figure.

(iii) Besides the non-spin-flip channel, the most important
spin-flip channels would be the $\mu =\pm 2$ channels if $g_{0}$ is
negative. However, if $g_{0}$ is positive, $\mu =0$ channel would be
a little more important, while $\sigma _{\pm 2}$ is the smallest. In
particular, when $g_0\approx g_6/2$, $\sigma _{\pm 2}$ is close to
zero.

The features of Fig.4 are quite popular. When different potentials
with different parameters are used, the qualitative features remain
unchanged.

\begin{figure}[tbp]
\centering \resizebox{0.8\columnwidth}{!}{
\includegraphics{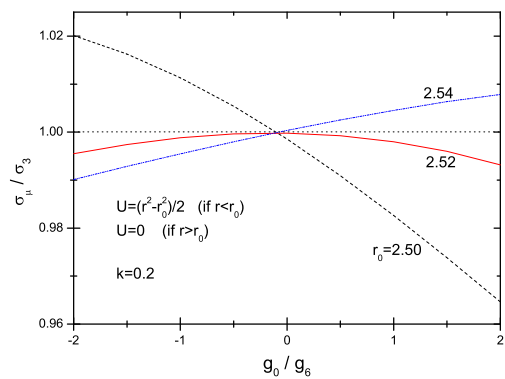} } \caption{(color online), The same as
Fig.3a but with $k=0.2$. }
\end{figure}

In previous examples $k=0.1$ is taken. When $k$ varies, the above
qualitative features remain. This is shown in Fig.5 to be compared
with Fig.3a. Note that, when $k$ varies, the optimal parameters of
the potentials would vary accordingly.

In summary, a specific s-wave scattering of a pair of $^{52}$Cr
atoms has been designed and studied theoretically.  Different from
the design of \cite{r_WJ2005}, both atoms are polarized but in
reverse directions previously and no magnetic field is applied. The
effect of $g_{0}$ on the scattering amplitudes without and with
spin-flip has been predicted. Whether $g_{0}$ is smaller or larger
than $g_{2}$ is found to be crucial to the qualitative features. In
order to determine the unknown $g_{0}$, associated experiments are
desirable.

\bigskip

\begin{acknowledgments}
The support from the NSFC under the grants 10574163, 10874249, and
from the project of National Basic Research Program of China
(2007CB935500) is appreciated.
\end{acknowledgments}

\bigskip

\appendix

\section{Matrix elements of the dipole-dipole interaction}

In this appendix the matrix elements of the dipole-dipole
interaction between the total spin-states $(12)_{SM}$ of a pair of
spin-3 atoms are given.

Making use of the spherical components of the spin-operator
$\mathbf{F}_{1}$ and $\mathbf{F}_{2}$, we have
\begin{eqnarray}
 &\langle(12)_{S'M'}&|V_{dd}|(12)_{SM}\rangle
 = -\frac{C_{d}}{r^{3}} 252\sqrt{4\pi }
   C_{1,0,\ 1,0}^{2,0} \sqrt{2S'+1} \nonumber \\
 &&\left\{
    \begin{array}{c c c}
     1 & 1 & 2 \\
     3 & 3 & S \\
     3 & 3 & S'
    \end{array}
   \right\}
   C_{2,M-M',\ S',M'}^{S,M} Y_{2,M-M'}(\hat{\mathbf{r}})
\end{eqnarray}
where the Clebsch-Gordan and 9-j symbols \cite{r_EAR1957} are
introduced, and $S+S^{\prime }$ \ must be even. From this formula it
is clear that $V_{dd}$ must cause a spatial relative rotation in
d-wave together with a change of $S$ (the choices of $S^{\prime }$
is $|S-2|$, $S$, and $S+2$).

\end{document}